\documentclass{aa}  
\usepackage{amssym,graphicx}
\begin{document}
\thesaurus{09                   
             (09.07.1;          
              09.18.1;          
              09.04.1;          
              09.08.1;          
              09.16.1;          
              11.09.1)}         

\title{\textit{Research note}\\
       Observational constraints on the ERE interpretation
       \thanks{Partly based on observations made at Observatoire de Haute Provence du
               CNRS and at European Southern Observatory (ESO), La Silla (Chile)}} 
\offprints{J.-P. Sivan}
\author{S.~Darbon, J.-M.~Perrin, J.-P.~Sivan}
\institute{Observatoire de Haute Provence du CNRS, F-04870 Saint Michel
l'Observatoire, France}
\date{Received xxx ; accepted yyy}
\maketitle

\begin{abstract}
Empirical relationships on the properties of the Extended Red Emission (ERE) are
presented. They are based on published observational data and on new results
obtained on reflection nebulae illuminated by cold stars. The plot of the width
versus the central wavelength of the ERE band is in agreement with laboratory
properties of the materials commonly proposed as the ERE carriers. But this is not the 
case for the plot of the ERE band width versus the effective temperature of the
nebula illuminating star.
\end{abstract}

\begin{figure}[!ht]
  \resizebox{\hsize}{8.5cm}{\includegraphics{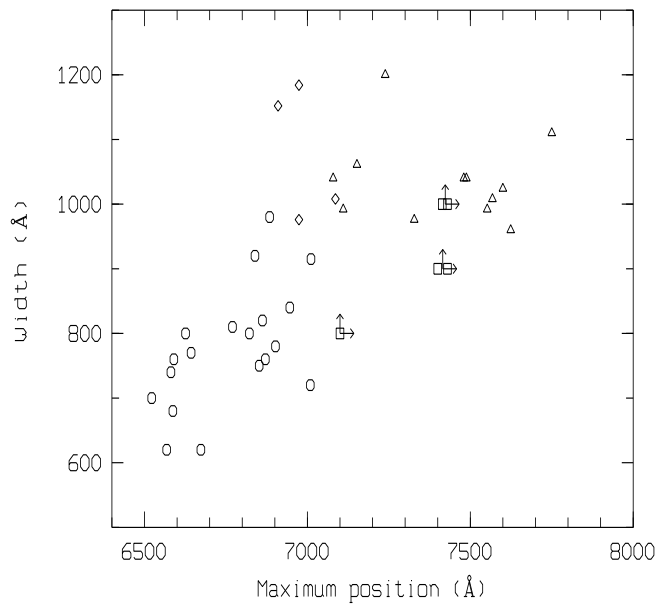}}
  \caption{This diagram plots the position of the maximum (\AA)
           versus the width (\AA) of the ERE band for a number of objects of various types :
           \ion{H}{ii}~regions ($\vartriangle$)(Perrin \& Sivan 1992 ; Sivan \& Perrin 
           1993 ; Darbon, Perrin \& Sivan 1998, Darbon et al., 1999), planetary nebulae 
           ($\square$)(Furton \& Witt 1992), reflection nebulae ($\circ$)(Witt \& 
           Boroson 1990) and the halo of M82 ($\lozenge$)(Perrin, Darbon \& Sivan 1995).
  \label{fig:eremaxwid}}
\end{figure}

\noindent
The Extended Red Emission (ERE) is a continuous emission band observed in the 
red part (6000--8000~\AA) of the spectrum of various astrophysical objects : 
reflection nebulae (Schmidt, Cohen \& Margon 1980; Witt \& Boroson 1990), 
planetary nebulae (Furton \& Witt 1992), galactic and extragalactic \ion{H}{ii} 
regions (Perrin \& Sivan 1992; Sivan \& Perrin 1993; Darbon, Perrin \& Sivan 1998), 
high-latitude galactic cirrus clouds (Szomoru \& Guhathakurta 1998), the halo of the 
galaxy M82 (Perrin, Darbon \& Sivan 1995) and the diffuse galactic medium (Gordon, 
Witt \& Friedmann 1998). The ERE band is found to vary significantly both in
position and width from an object to an other. The diversity of objects where
the ERE is detected and the diversity of the observed band characteristics lead us
to search for empirical relationships that should help identification of the ERE
carriers.\\
For all the available observations, we have plotted in Fig.
\ref{fig:eremaxwid} the width as a function of the central wavelength of the
ERE bands. For reflection nebulae, we have used the values of Witt \& Boroson
(1990). For all the other objects, we have made measurements according to the
same definitions as Witt \& Boroson : the central wavelength splits the band luminosity 
into two equal parts and the width is measured as the difference between the wavelengths
of the first and third quartiles. Note that for planetary nebulae the values
derived from the published spectra of Furton \& Witt (1992) are
underestimated.\\

The diagram in Fig. \ref{fig:eremaxwid} reveals a clear correlation between the 
position of the maximum and the width of the band. We derived a correlation
coefficient $r=0.68$. This result confirms and reinforces the tendency 
previously noted by Witt \& Boroson (1990) from reflection nebulae only : the
correlation coefficient was $r=0.52$. The same effect 
is observed in laboratory experiments : Hydrogenated Amorphous Carbon (HAC) grains
(Furton \& Witt 1993) and nanocrystals of silicon (Witt, Gordon \& Furton 1998 ; Ledoux et 
al. 1998) exhibit luminescence bands whose width increases with the maximum wavelength.\\
Also, it is found that the plotted values in Fig. \ref{fig:eremaxwid} are split
into two groups : reflection nebulae with smaller widths and bluer peaks and
\ion{H}{ii} regions and planetary nebulae with larger widths and redder peaks
(the halo of M82 lies at the border of these two groups). This repartition might be
related to the presence or absence of plasma within the nebula, in agreement
with laboratory results (Wagner \& Lautenschlager 1986; Robertson \& O'Reilly
1987)

\begin{figure}[!ht]
  \resizebox{\hsize}{8.5cm}{\includegraphics{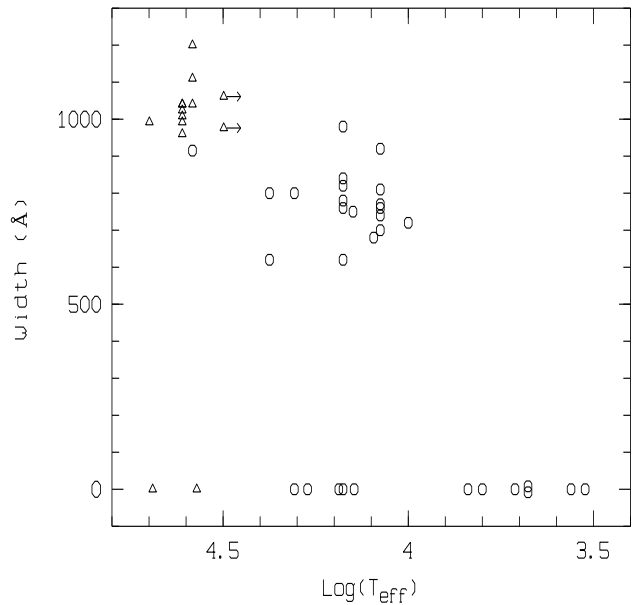}}
  \caption{Width of the ERE band as a function of the logarithm of the effective
           temperature of the exciting and/or illuminating star as measured on
           \ion{H}{ii} regions ($\vartriangle$) and reflection nebulae ($\circ$). 
           The exciting star of Sh 152 is embedded in a dusty cocoon so that the
           effective temperature of the spectral energy distribution that
           actually illuminates the regions observed by Darbon et al. (1999) is 
           over-estimated. 
  \label{fig:ereteff}}
\end{figure}

As a mean, in Fig. \ref{fig:eremaxwid}, the reflection nebulae are illuminated by
stars less energetic than the other nebulae. This repartition suggests
that the characteristics of the ERE might depend on the spectral distribution
of the exciting flux. This has lead us to study the variation of the ERE
band width as a function of the effective temperature of the exciting stars. 
BVI photometric and spectrophotometric observations, conducted respectively by Witt \& Schild
(1985) and Witt \& Boroson (1990), deal with reflection nebulae illuminated by stars with a
spectral type earlier than A0 (i.e. T$_{eff}\gtrsim 10000$~K) : most of these nebulae exhibit 
ERE. But laboratory results show that the ERE can be excited by low energy visible 
radiation. So, the observation of reflection nebulae illuminated by stars colder than A0 (i.e. with a 
spectral energy distribution peaking in the visible) should be useful in order to detect the 
presence or the absence of the ERE in their spectra. One such nebula, illuminated by an M 
giant star, has been observed by Witt \& Rogers (1991). We have observed six additional
nebulae illuminated by cold stars. They are listed in Table \ref{tbl:rncold}, together with 
spectral type and the effective temperature of the illuminating stars. We obtained low
resolution spectra for these objects. Data reduction were conducted as previously described in 
Perrin, Darbon \& Sivan (1995). None of these nebulae exhibits ERE in its spectrum.

\begin{table}[!ht]
\begin{center}
\caption{\label{tbl:rncold} 
Reflection nebulae illuminated by cold stars}  
\begin{flushleft}
\begin{tabular}{lccc} 
\hline
\hline 
Nebula    & Spectral Type (MK) & T$_{eff}$ (K) & Observation site \\
\hline
VDB003                     & K0III    & 4750 & OHP (a) \\
VDB035                     & G5       & 5150 & OHP (a) \\
VDB037                     & (g)M5III & 3330 & OHP (a) \\
VDB120                     & F7II     & 6310 & OHP (a) \\
VDB133$^{\dagger}$         & F5Iab    & 6900 & OHP (a) \\
VHE14B                     & K0III    & 4750 & ESO (b) \\
\hspace{0.5mm}IC2220       & (g)M1II  & 3635 & UTSO (c)\\
\hline
\end{tabular}
\end{flushleft}
\end{center}
\begin{list}{}{}
\item[(a)] observed at the Cassegrain focus of the 193 cm telescope of the Observatoire 
           de Haute Provence du CNRS using the Carelec spectrograph equipped with a
           512$\times 512$ thinned back-illuminated Tektronic CCD 
\item[(b)] observed at the Cassegrain focus of the 152 cm telescope of the European 
           Southern Observatory at La Silla using the Boller \& Chivens spectrograph equipped 
           with a 2048$\times 1024$ thinned back-illuminated CCD
\item[(c)] observed by Witt \& Rogers (1991) with the University of Toronto 
           Southern Observatory (UTSO) 61-cm telescope equipped with a CCD 
           camera and B, V, R, I filters.
\item[$^{\dagger}$] the main illuminating star HD195593A has a hotter companion
                    HD195593B (spectral type B6-B8) (Uchida, Sellgren \& Werner 1998).
\end{list}
\end{table}

\begin{figure}[!ht]
  \resizebox{\hsize}{8.5cm}{\includegraphics{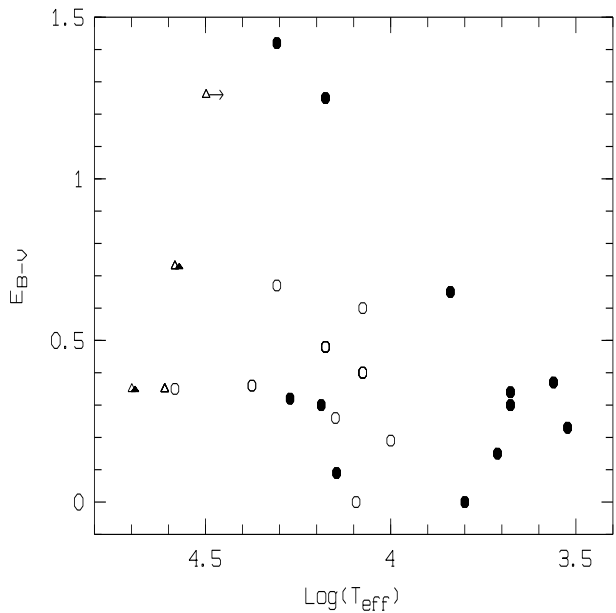}}
  \caption{Color excess as a function of the logarithm of the effective
           temperature of the illuminating star for : 
           \ion{H}{ii} regions with ERE ($\vartriangle$), reflection
           nebulae with ERE ($\circ$), \ion{H}{ii} regions without ERE ($\blacktriangle$) 
           and reflection nebulae without ERE ($\large\bullet$).  
  \label{fig:ereebv}}
\end{figure}

In Fig. \ref{fig:ereteff}, we have plotted the ERE band width as a function of the
effective temperature of the exciting star for the reflection nebulae and \ion{H}{ii} 
regions of Fig. \ref{fig:eremaxwid} and for the nebulae of Table \ref{tbl:rncold}. For 
the nebulae without ERE, the width value is set to zero.
Clearly, no ERE appears for T$_{eff}\lesssim 7000$~K. On the contrary, for
T$_{eff}\gtrsim 10000$~K, most of the plotted objects exhibit the ERE albeit few of
them do not. This suggests a cut-off in effective temperature might exist
between 7000 and 10000 K (unfortunately no observation is available for this
range). This cut-off cannot be accounted for by extinction effects as
demonstrated by Fig. \ref{fig:ereebv} : the presence or the absence of the ERE
in a nebula does not appear to be correlated to the color excess of its
illuminating star. \\
The fact that no ERE is present for stars whose 
effective temperature is smaller than 7000 K does not agree with current
laboratory data on various materials such as HACs and silicon nanocrystals. These materials
exhibit red luminescence features when they are irradiated by photons of low energy, namely of
energy smaller than 3 eV (see e.g. Sussmann \& Ogden 1980, Lin \& Feldman 1981, Fang et al. 1988, 
Wilson et al. 1993).\\
Further observations of reflection nebulae illuminated by cold stars would be
useful to confirm our finding. We cannot completely rule out that the absence of ERE in 
these nebulae could be simply due to the absence of any luminescent material.\\ 
This is probably the case for the nebulae without ERE belonging to the left-hand part of the 
diagram in Fig. \ref{fig:ereteff} (T$_{eff}\gtrsim 10000$~K). This suggests that the ERE 
carriers would not be present everywhere in the interstellar medium. 

\end{document}